# Novel nanometer-level uniform amorphous carbon coating for boron powders by direct pyrolysis of coronene without solvent

ShuJun Ye, MingHui Song, and Hiroaki Kumakura

National Institute for Materials Science, Sengen 1-2-1, Tsukuba, 305-0047, Japan

Corresponding E-mail： YE.Shujun@nims.go.jp and KUMAKURA.Hiroaki@nims.go.jp

# Notice

**Abstract**

A 3 nm coronene coating and a 4 nm amorphous carbon coating with uniform shell-core encapsulation structure for nanosized boron (B) powders are formed by a simple process, where coronene is directly mixing with boron particles without a solvent, and heated at 520 °C for 1 h or at 630 °C for 3 h in a vacuum-sealed silica tube. Coronene has a melting point lower than its decomposition temperature, which enables liquid coronene to cover B particles by liquid diffusion and penetration without the need for a solvent. The diffusion and penetration of coronene can extend to the boundaries of particles and inside agglomerated nanoparticles to form a complete shell-core encapsulated structure. As the temperature is increased, thermal decomposition of coronene on the B particles results in the formation of a uniform amorphous carbon coating layer. This novel and simple nanometer-level uniform amorphous carbon coating method is possibly applied to many other powders, thus it has potential applications in many fields with low cost.

**Introduction**

Amorphous carbon coatings have many applications in materials science due to characteristics such as low friction, high hardness, high elastic modulus, chemistry inertness, biocompatibility, high moisture resistance, and high wear resistance [1-8]. Amorphous carbon coating can also be used to modify Li-ion battery anode materials (e.g., $SnO_2$, $LiVPO_4F$, $LiFePO_4$, $Li_2S$, and $SiO_2$) [9-18] because it is particularly effective to improve conductivity and surface chemistry. The applications of amorphous carbon coating on photocatalysts such as $TiO_2$ [19-20] and on capacitors electrode materials [21-22] have also been reported.

Amorphous carbon coatings can be produced by a wide range of methods such as ion beam deposition, plasma decomposition, magnetron sputtering, ion beam sputtering, laser plasma deposition, chemical vapor deposition (CVD), and plasma-enhanced CVD [1-4]. However, all of these methods are high cost and involve complex processes. In addition, it is difficult to obtain uniform carbon coating layer with some of these sputtering and deposition methods [7-9]. Pyrolysis of organic materials is also used to produce carbon coatings, where the organic materials are typically dissolved in a solvent to make a uniform mixture and the target material is coated with the mixture, dried and then thermally decomposed. In some cases, the pH and temperature must be adjusted to improve the solubility in the solvent, and the products need to be washed to decrease impurities. Although some methods have achieved a uniform carbon coating layer, it is more difficult to obtain uniform carbon coating by pyrolysis than by the other methods. Furthermore, the thermal decomposition products from the solvent, whether water or an organic solvent is used, result in impurities that are unfavorable for some applications. There have been very few reports on the pyrolysis of organic precursors without the use of a solvent. Tsumura et al. [19] used poly (vinyl alcohol) as a carbon source that was directly mixed with nanosized $TiO_2$ and heated the mixture at 700-1000 °C for 1 h under nitrogen gas flow (100 mL/min). However, slight sintering and crystal growth of the $TiO_2$ were evident in the carbon-coated $TiO_2$ product. These results indicate that the $TiO_2$ surface is not uniformly coated with the poly (vinyl alcohol) by direct mixing without a solvent. Some materials are incompatible with the solvent. In addition, some nanoparticles are agglomerated and if the nanoparticle boundaries and inside of agglomerated nanoparticles are considered, then it is difficult to obtain a uniform coating layer (shell-core encapsulation) even with the expensive sputtering and deposition methods.

In this study, we report a novel and simple method to achieve uniform (shell-core encapsulation) amorphous nanometer-level carbon coating by direct mixing and pyrolysis of coronene (carbon source) without a solvent. Nanosized boron (B) particles were used as the target material for carbon coating.

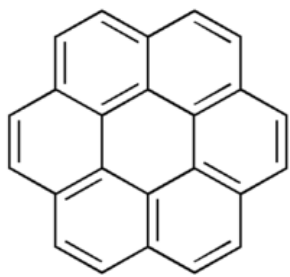

Fig. 1. Molecular structure of coronene ($C_{24}H_{12}$).

Coronene ($C_{24}H_{12}$; C24) is a yellow powder with a melting point of 438 °C and a boiling point of 525 °C. It is an aromatic hydrocarbon comprising six peri-fused benzene rings, of which the structure is shown in Fig. 1. The structure of six peri-fused benzene rings has the highest symmetry among all organic compounds except benzene. Talyzin et al. systemically studied the characteristics of C24 during a heat treatment process [23], where its dimer dicoronylene was obtained at 530-550 °C, followed by oligomerization into larger molecules at 550-600 °C, and then thermal decomposition of the large molecules into graphitic nanoparticles above 600 °C. These thermal characteristics of C24 led us to consider it as a carbon source for carbon coating without a solvent in this study.

**Experimental details**

B powder (amorphous, 250 nm average particle size, 98.5% purity) and 0.29 mol% C24 (7mol% carbon to boron) with a purity of 95% were uniformly mixed and ground in a mortar, and then packed in a silica tube (0.2 cm thick, 3 cm O.D, 28 cm long) under vacuum. Two samples (C24+PB@520-1h and C24+PB@630-3h) were prepared as shown in Fig. 2, where C24+PB@520-1h was 0.7 g of mixed powders that was heat treated at 520 °C for 1 h, and C24+PB@630-3h was 2.7 g of mixed powders that was heat treated at 630 °C for 3 h. The C24+PB@630-3h powder was prepared with a larger weight for use in other applications. Each powder was placed into one end of the silica tube and set in a uniform temperature region (about 10 cm) of the furnace during the heat treatment; the other end of the silica tube was in a temperature gradient and the head of the silica tube was at the lowest temperature.

X-ray diffraction (XRD) patterns were measured with a Rigaku Miniflex II diffractometer. High resolution transmission electron microscopy (TEM) observations and electron energy loss spectroscopy (EELS) were conducted using a Tecnai G2 F30 microscope. Samples for TEM were prepared as follows. A small target particle was placed into 1 mL deionized water and ultrasonically dispersed for 3 min, and then 0.5 mL of the dispersion was added to 2 mL deionized water further ultrasonically dispersed for 3 min. One droplet of the final dispersion was dropped onto a Cu microgrid and dried at 30 °C in air for 2 h. Infrared (IR) absorption spectroscopy was conducted with a Spectrum GX-Raman spectrometer. 0.01 g of sample powder was uniformly mixed and ground with 0.18 g KBr in a mortar, and then pressed into a pellet for IR measurement.

**Results**

 Fig. 2(a) (C24+PB@520-1h) shows a yellow powder at the head part of the silica tube. C24 has a boiling point of 525 °C under ambient pressure; therefore, sublimated C24 gas should be present at a temperature of 520 °C, and the yellow powder (C24) precipitates at the head region of the silica tube when cooling to room temperature (RT). Talyzin et al. [23] reported that C24 is partially transformed into dicoronylene at 500-530 °C; however, the ruby red colored dicoronylene was not observed. This may be due to the smaller amount of C24 used in the present study compared to that used by Talyzin et al., so that any dicoronylene formed would be too small to be observed. Talyzin et al. [23] also reported that C24 is thermally decomposed to graphitic nanoparticles above 600 °C. Fig. 2(b) (C24+PB@630-3h) was heated at 630 °C for 3 h; therefore, the black powders adsorbed to the inner wall of the silica tube are considered

to be graphitic nanoparticles.

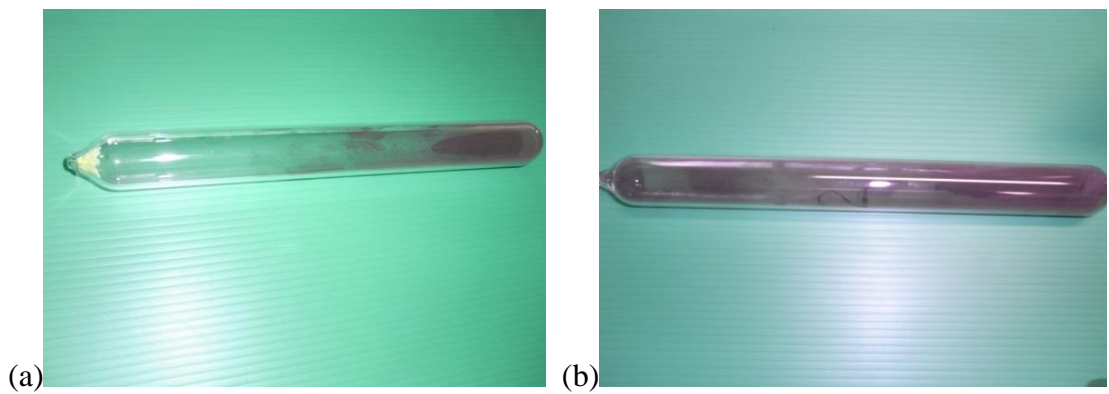

Fig. 2. Photographs of samples (a) C24+PB@520-1h, and (b) C24+PB@630-3h.

Fig. 3 shows XRD patterns for pure C24, pure B (PB), C24+PB at RT, and samples C24+PB@520-1h and C24+PB@630-3h. C24 has a strong 200 peak at 11.8° and many other peaks. PB has three broad amorphous peaks (center positions at 22.06°, 35.5°, and 68.4°), and two small peaks at 14.5° and 27.7° due to crystalline boron. The XRD pattern for the powders ground in a mortar without heat treatment (C24+PB@RT) shows the accumulated peaks of both PB and C24. The peaks are weaker than that for the pure C24 powder because there is only 0.29 mol% C24 to B in C24+PB @ RT. Sample C24+PB@520-1h, taken from the end part of silica tube, has weaker C24 peaks than C24+PB@RT. Sample C24+PB@630-3h heat treated at 630 °C for 3 h shows only broad amorphous peaks of B, which suggests that C24 has been completely decomposed.

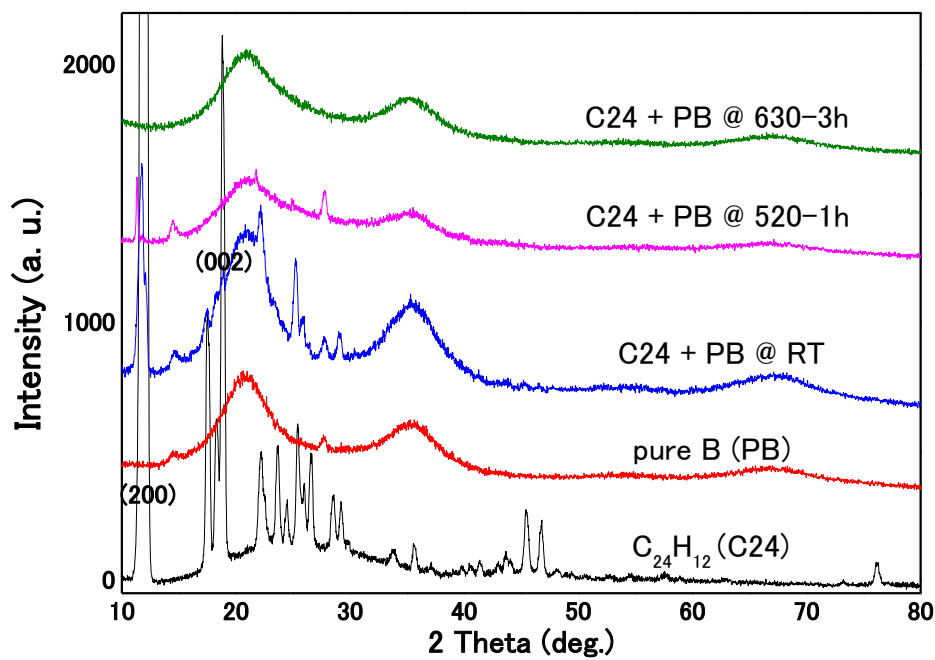

Fig. 3. Comparison of XRD patterns for all powders.

Fig. 4 shows TEM images of the C24+PB@630-3h sample. Fig. 4(a) shows a cluster of dispersed powder on the Cu microgrid. Fig. 4(b) shows a magnified image of the region indicated in red in Fig. 4(a). Thin coating layers can be observed in all the particles shown in Fig. 4(b). The region marked in red in Fig. 4(b) is further magnified and shown in Fig. 4(c) (counterclockwise rotation of 90°). A uniform coating layer with a thickness of approximately 4 nm can be observed around all the particles. It should be noted that the coating layer is also present at the boundary between two particles. Fig.

4(d) shows an EELS spectrum obtained from the area of Fig. 4(c). The energy loss edges of C-K at about 284 eV and that of B-K at about 188 eV can be clearly observed, which confirms the presence of carbon on the boron particles.

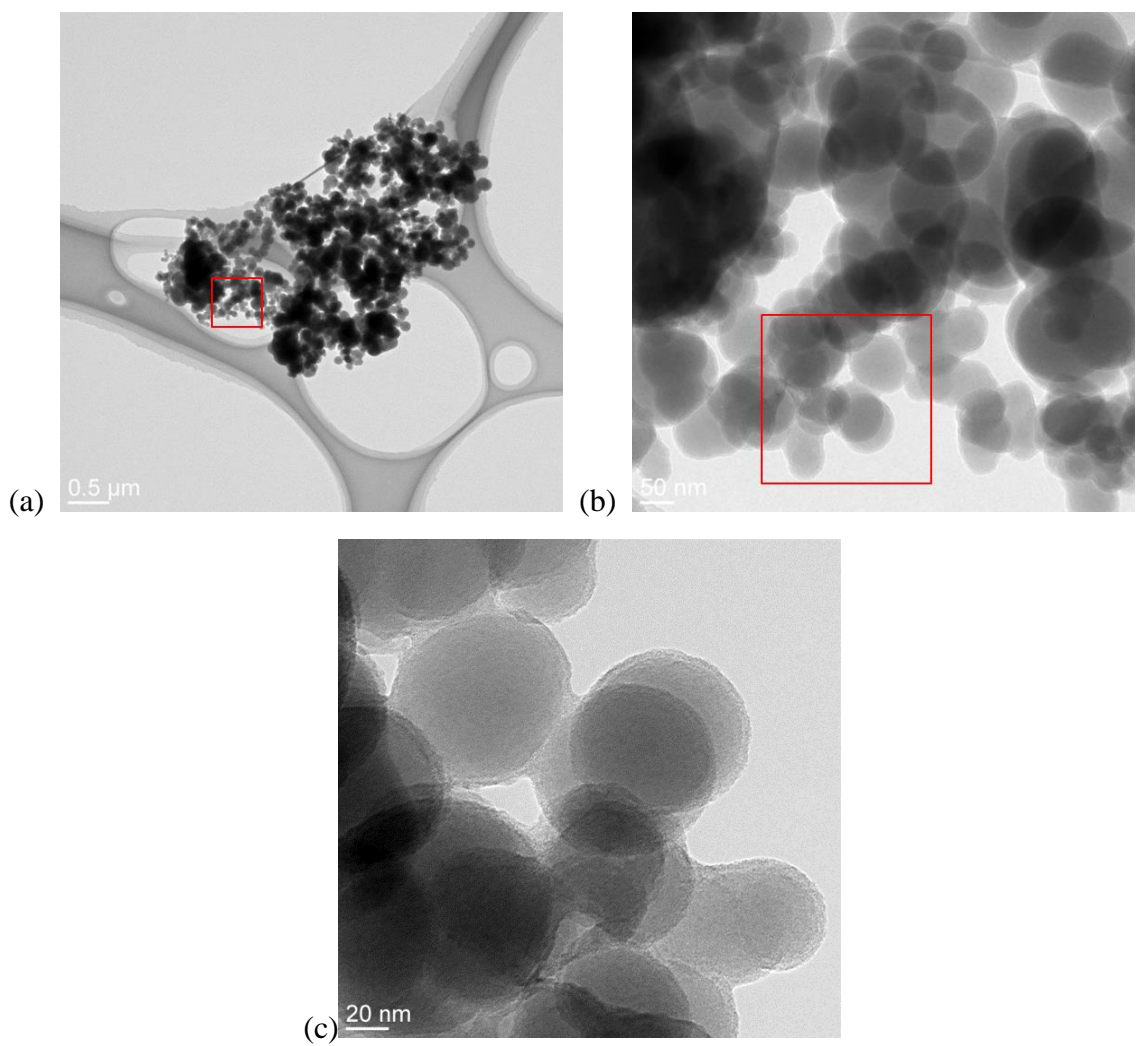

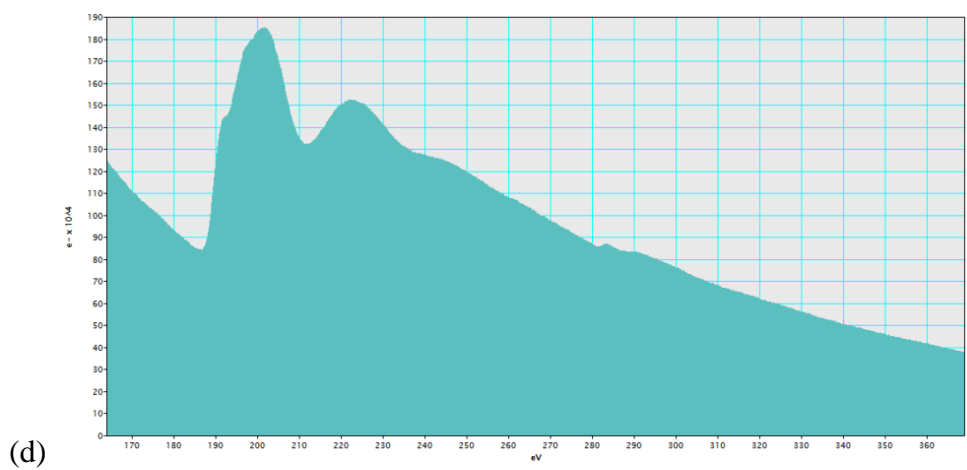

(d)

Fig. 4. TEM images of (a) dispersed C24+PB@630-3h, (b) magnified image of region marked with red in (a), (c) magnified image of region marked with red in (b) (counter clockwise rotation of 90°), and (d) EELS spectrum obtained from the area shown in (c).

To determine the distribution of carbon on the boron particles, EELS elemental mappings of C and B were obtained with C-K core energy loss electrons and B-K core energy loss electrons, respectively. A TEM image of a cluster of particles is shown in Fig. 5(a). Boron and carbon maps of Fig. 5(a) are shown in Figs. 5(b) and (c), respectively. Fig. 5(d) shows a superimposed image of Figs. 5(b) and (c), which illustrates the distribution of both C and B. A uniform carbon coating with a thickness of about 4 nm can be observed on the surface of all the small particles in the upper region of the figure. The lower region of the figure has larger particles and/or overlapped particles that is too thick to give a sufficiently strong relative contrast of carbon out of the background. These results and the identification of a carbon layer at the boundary of two particles shown in Fig. 4(c) indicates that C24 and/or larger hydrocarbon molecules are still present on the surface of the born particles after heat treatment above the boiling point of C24 and decomposition to form an amorphous carbon layer at 630 °C. High-resolution TEM images of the red region indicated in Fig. 5(a) are shown in Figs. 5(e) and (f). Here, both boron and carbon are in the amorphous state, which is consistent with the XRD pattern shown in Fig. 3.

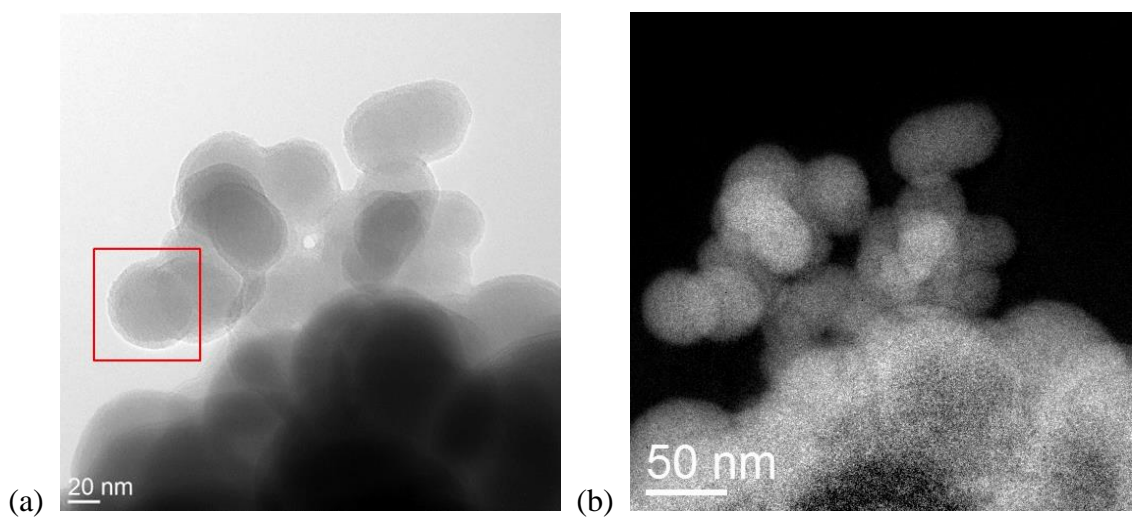

(a)  (b)

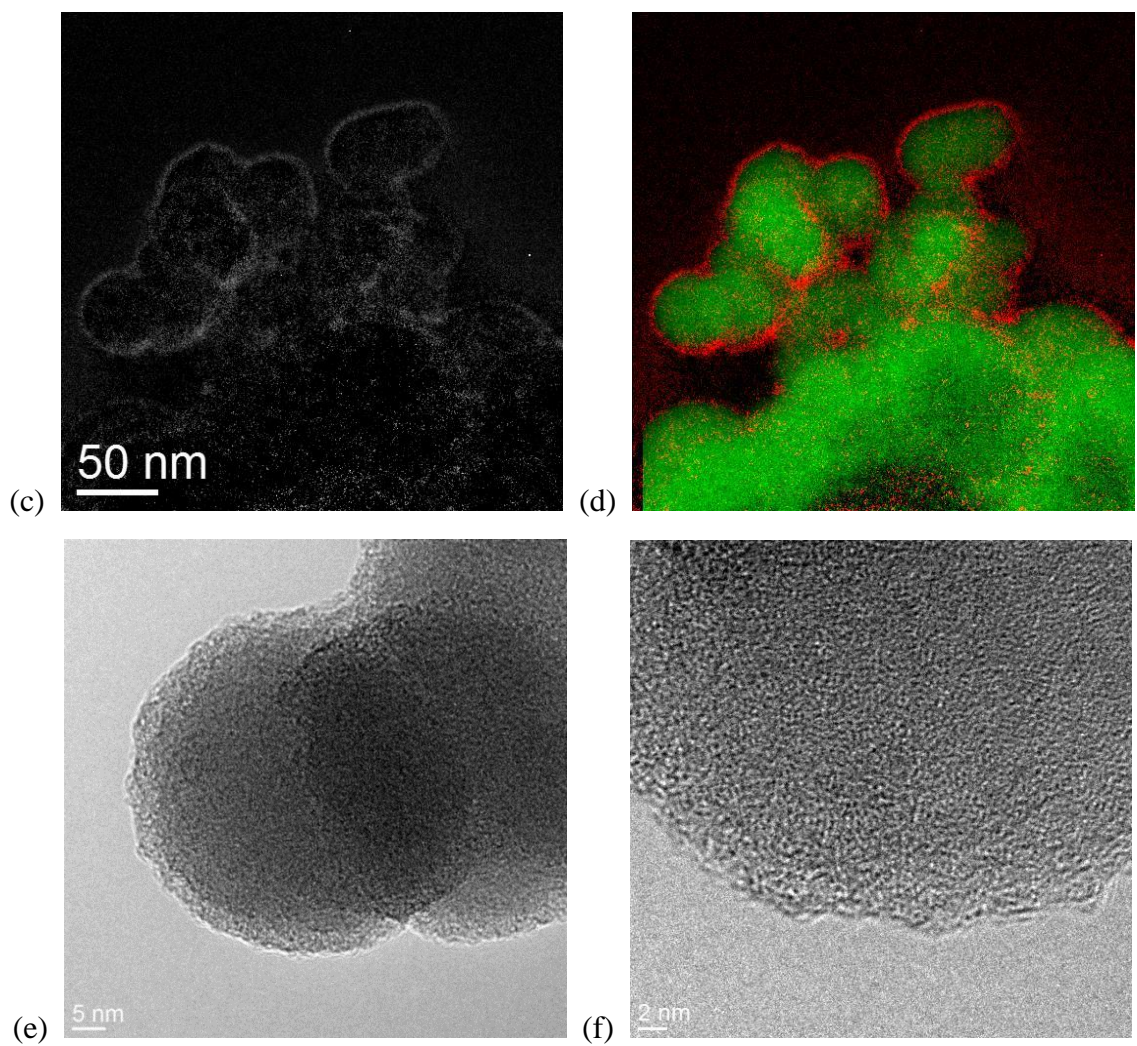

Fig. 5. (a) TEM image of the C24+PB@630-3h powder. EELS elemental maps for (b) B and (c) C, and (d) superimposition image of (b) and (c). (e) High-resolution image of the red region marked in (a), and (f) magnified image of (e).

Figs. 6(a) and (b) show TEM images of PB and C24+PB@520-1h, respectively. PB has no coating layer, while C24+PB @ 520-1h has uniform coating layers with a thickness of about 3 nm on the B particles. Elemental mapping and EELS also indicate that these coating layers contain carbon. Combined with the XRD results, these coating layers are attributed to C24 coating. C24, when heat treated at 520 °C (higher than its melting point and lower than its boiling point) can diffuse to the entire surface of the B powder. C24 can even diffuse to the boundaries and permeate to the inside of agglomerated nanoparticles. The large surface area of B nanoparticles and its large absorption capacity support the diffusion and penetration of C24 to form a C24 coating on B particles.

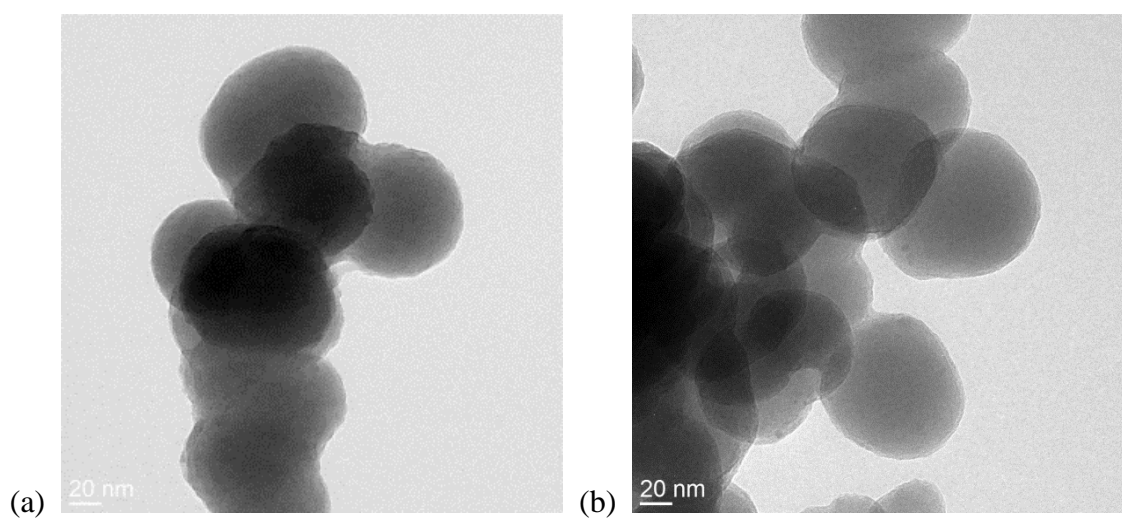

Fig. 6. TEM images of (a) PB and (b) C24+PB@520-1h powders.

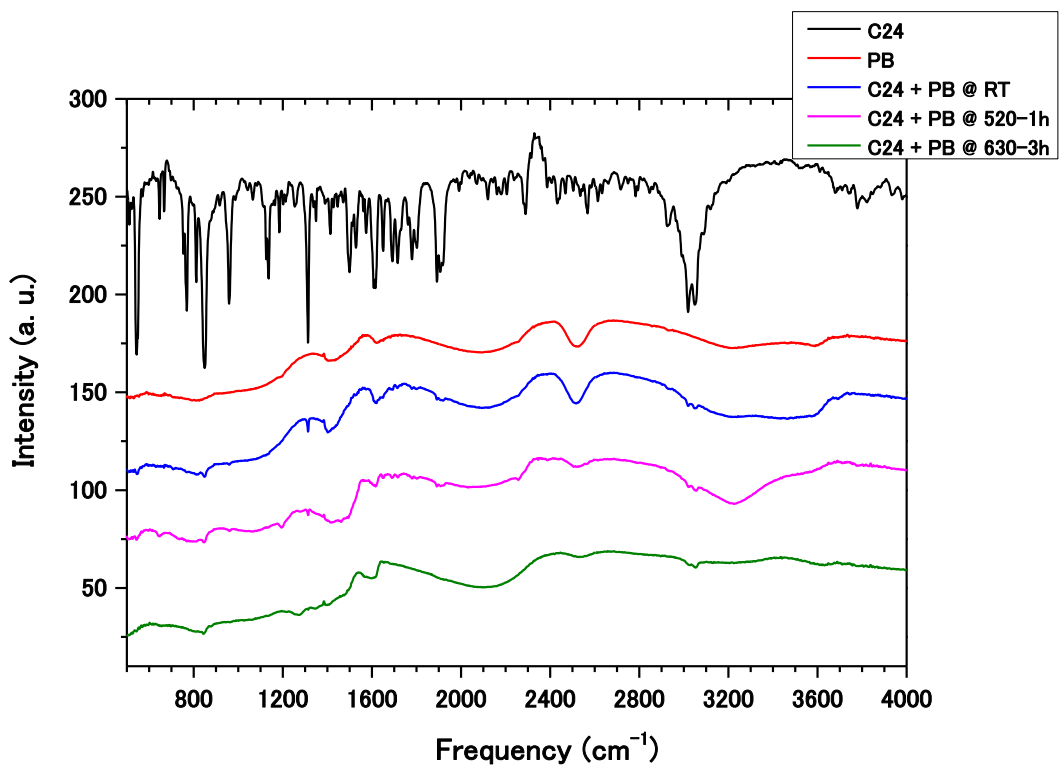

Fig. 7. Comparison of IR absorption spectra for all powders.

Fig. 7 shows a comparison of the IR spectra for the five powders. The spectrum for C24+PB@RT showed the accumulated peaks of PB and C24. C24+PB@RT is the PB with only 0.29 mol% C24, thus C24 peaks of C24+PB@RT are much weaker than that for the pure C24 powder. The peaks attributed to C24 obtained by Colangeli et al. [24] can be clearly observed in C24+PB@RT; aromatic C-H stretching at 3020 and 3050 cm$^{-1}$, overtones and/or combinations of fundamental modes at 1920, 1892, 1800, 1780, 1715, and 1692 cm$^{-1}$, aromatic C-C stretching at 1605 and 1317 cm$^{-1}$, C-H out-of-plane bending at 957 and 857 cm$^{-1}$, and C-C-C out-of-plane bending at 544 cm$^{-1}$. All the peaks show decrease of intensity in the order C24+PB@RT > C24+PB@520-1h > C24+PB@630-3h. In particular, the peaks of overtones and/or combinations of fundamental modes have disappeared in the C24+PB@630-3h spectrum, which suggests that C24 is completely thermally decomposed. The IR spectrum results, consistent with the XRD pattern in Fig. 3 and TEM images in Figs. 4 and 5, proves that the coating layer for C24+PB@630-3h are amorphous carbon.

**Discussion**

The diffusion and penetration of C24 to form a uniform coating structure can only be obtained with the liquid state. Gas and vapor caused during coating with sputtering and deposition methods can be adsorbed on target particles to obtain a coating; however, diffusion and penetration to the inside and boundaries of agglomerated particles cannot occur. Therefore, the present method is the only method that can realize a uniform carbon coating (shell-core encapsulation structure) for particles without the use of a solvent. Solvents tend to introduce impurities; therefore, the proposed C24 coating method, after the thermal decomposition of C24, is the only method to produce a uniform carbon coating layer on nanoparticles without the introduction of impurities.

This carbon coating formed by C24 coating is achieved by a simple and low-cost process, which is thus expected to be used for large-scale applications.

The thickness of the C24+PB@520-1h coating layer is about 3 nm, which is less than that of the C24+PB@630-3h at about 4 nm. The reason for this is mainly the larger amount (almost 3.9 times) of C24+PB @ C24+PB@630-3h starting powders (2.7 g) sealed in the same size silica tube than that used to prepare C24+PB@520-1h (0.7 g). If a further heat treatment of the C24+PB@520-1h sample was conducted at 630 °C for 3 h, then the thickness of the carbon coating would be expected to be around 3 nm rather than 4 nm. This is because decomposition of C24 can`t change the thickness of coating layer too much due to small atomic percentage of hydrogen. This also suggests that the thickness of the carbon coating can be adjusted to satisfy different applications.

B powders were investigated here based on our previous research, which has been focused on $MgB_2$ superconducting wires [25]. Carbon doping on B powder can enhance the upper critical field and in-field critical current density ($J_c$) of $MgB_2$ superconducting wires [26]. The C24 coated boron powders were used to prepare $MgB_2$ wires and high $J_c$ values [25] were obtained, similar to those for $MgB_2$ wires fabricated with the carbon-doped B powders prepared by the RF plasma method [27-28]. Kim et al. [29] reported that carbon-doped B powders produced by RF plasma are carbon-encapsulated B powders. However, we consider that the synthesis process for carbon-doped B (reactions of mixing $BCl_3$, $H_2$, and $CH_4$ gases in RF plasma) [30] would not result in 100% carbon-encapsulation, but the formation of carbon-encapsulated boron may be dominant over boron-encapsulated carbon. In our previous research [25], we supposed that the surfaces of B particles would be coated with C24 after heating at 450 °C (slightly higher than the melting point of C24, 438 °C) for only 10 min in a sealed silica tube under vacuum, and then after high-density packing of the C24-coated B powders to produce $MgB_2$ wires, we considered this C24 coating would be equivalent to a carbon coating because C24 contains 96 wt% carbon and pyrolytically decomposes to graphitic nanoparticles above 600 °C (lower than the reaction temperature for the formation of $MgB_2$). After the investigation for many samples, it was determined that C24 covers the B particles even at a higher temperature of 520 °C (much higher than the melting point and slightly lower than the boiling point of C24, 525 °C), so that a carbon coating (C24+PB@630-3h) can be directly obtained even in a sealed vacuum silica tube (not a high-density packed space). Actually, we have obtained high $J_c$ values by using C24+PB@630-3h, which are similar to that obtained from the C24-coated [25] or RF plasma produced [25, 27-29] B powders. The $J_c$ values are expected to be further enhanced by adjustment of the thickness of carbon layer, which will be described in our late paper.

But in this paper, we just want to introduce this uniform amorphous carbon coating method and show its potential applications in many fields as introduced in introduction section.

Any particles, if they keep stable and do not react with coronene below the temperature of carbon coating formation (about 600 °C), could be coated with carbon by the method developed in this study. Thus, we consider that the carbon-coating method developed in this study could be used for Li-ion battery anode materials (e.g., $SnO_2$, $LiFePO_4$, $Li_2S$, and $SiO_2$) [9-18], photocatalysts (e.g., $TiO_2$ [19, 20]), capacitors electrode materials (e.g., $Fe_3O_4$ [21], Porous Silicon nanowires [22]), and so on.

Recently, coronene has been actively studied for the formation of films on plate substrates [31-32]. These coronene coatings on substrates could also be pyrolyzed to form amorphous carbon coatings. Coronene is also widely used for encapsulation of carbon nanotubes [33-35]. When these types of composites are heat treated, amorphous carbon-coated carbon nanotubes could be formed, which are also expected to find use in many new applications. In addition, not only coronene, but other polyaromatic hydrocarbons [36] may also be used as a carbon source for this new carbon coating method.

**Conclusion**

In this work, a simple process for uniform nanometer-level (shell-core encapsulation) carbon coating was developed by the direct pyrolysis of coronene on boron particles without a solvent. Coronene has a melting point lower than its decomposition temperature, which enables coronene to be coated on B particles by liquid diffusion to the particle boundaries and penetration to the inside of agglomerated nanoparticles to form a completely shell-core encapsulated structure. As the temperature is increased, the coated coronene is thermally decomposed to form a uniform amorphous carbon coating (shell-core encapsulation) without the need for a solvent. This simple and low-cost process of nanometer-level uniform amorphous carbon coating is expected to have potential applications in many fields.

**Acknowledgements**

The authors are grateful to Mr. H. Takigawa, Drs. H. Kitaguchi, K. Togano, A. Matsumoto, H. Fujii, and G. Nishijima and all other staff members in the superconducting wire unit of the National Institute for Material Science (NIMS) for their kind assistance and for useful discussions. S. J. Ye also thanks Misses N. Isaka, Y.

Nishimiya, K. Nakayashiki, A. Shichiri, J. Li, Drs. T. Minowa, and M. Takeguchi in NIMS for the kind help during the experiments. This study was supported by the NIMS Molecule & Material Synthesis Platform in the "Nanotechnology Platform Project" operated by the Ministry of Education, Culture, Sports, Science and Technology (MEXT) of Japan. Financial support for this work was provided by the Advanced Low Carbon Technology Research and Development Program (ALCA) of the Japan Science and Technology Agency (JST).